\documentclass[aps,prl,twocolumn,amsmath,amssymb,nofootinbib,superscriptaddress,floatfix,reprint,longbibliography]{revtex4-1}

\usepackage{graphicx}
\usepackage{dcolumn}
\usepackage{bm}
\usepackage{tabularx}
\usepackage{amsmath}
\usepackage{textcomp}
\usepackage{upgreek}
\makeatletter

\newcommand{\Rmnum}[1]{\expandafter\@slowromancap\romannumeral #1@}
\makeatother

\usepackage[colorlinks=true, urlcolor=blue, linkcolor=blue, citecolor=blue, pdftex]{hyperref}
\usepackage{ulem} 
\usepackage[pdftex]{color} 

\begin{document}

\title{Competition of superconductivity and charge density wave in selective oxidized CsV$_3$Sb$_5$ thin flakes}

\author{Yanpeng Song$^\dagger$} \affiliation{Beijing National Laboratory for Condensed Matter Physics, Institute of Physics, Chinese Academy of Sciences, Beijing 100190, China}
\author{Tianping Ying$^\dagger$} \affiliation{Beijing National Laboratory for Condensed Matter Physics, Institute of Physics, Chinese Academy of Sciences, Beijing 100190, China} \affiliation{Materials Research Centre for Element Strategy, Tokyo Institute of Technology, Yokohama 226-8503, Japan}
\author{Xu Chen}\affiliation{Beijing National Laboratory for Condensed Matter Physics, Institute of Physics, Chinese Academy of Sciences, Beijing 100190, China}
\author{Xu Han}\affiliation{Beijing National Laboratory for Condensed Matter Physics, Institute of Physics, Chinese Academy of Sciences, Beijing 100190, China}
\author{Xianxin Wu$^*$}\affiliation{Max-Planck-Institut f\"ur Festk\"orperforschung, Heisenbergstrasse 1, D-70569 Stuttgart, Germany}
\author{Andreas P. Schnyder}\affiliation{Max-Planck-Institut f\"ur Festk\"orperforschung, Heisenbergstrasse 1, D-70569 Stuttgart, Germany}
\author{Yuan Huang}\affiliation{Beijing National Laboratory for Condensed Matter Physics, Institute of Physics, Chinese Academy of Sciences, Beijing 100190, China}
\author{Jiangang Guo$^*$}\affiliation{Beijing National Laboratory for Condensed Matter Physics, Institute of Physics, Chinese Academy of Sciences, Beijing 100190, China}\affiliation{Songshan Lake Materials Laboratory, Dongguan 523808, China}
\author{Xiaolong Chen$^*$}\affiliation{Beijing National Laboratory for Condensed Matter Physics, Institute of Physics, Chinese Academy of Sciences, Beijing 100190, China} \affiliation{Songshan Lake Materials Laboratory, Dongguan 523808, China}\affiliation{University of Chinese Academy of Sciences, Beijing 100049, China}

\date{\today}

\begin{abstract}
The recently discovered layered kagome metals AV$_3$Sb$_5$ (A = K, Rb, and Cs) with vanadium kagome networks provide a novel platform to explore correlated quantum states intertwined with topological band structures. Here we report the prominent effect of hole doping on both superconductivity and charge density wave (CDW) order, achieved by selective oxidation of exfoliated thin flakes. A superconducting dome is revealed as a function of the effective doping content. The superconducting transition temperature ($T_{\mathrm{c}}$) and upper critical field in thin flakes are significantly enhanced compared with the bulk, which are accompanied by the suppression of CDW. Our detailed analyses establish the pivotal role of van Hove singularities (VHSs) in promoting correlated quantum orders in these kagome metals. Our experiment not only demonstrates the intriguing nature of superconducting and CDW orders, but also provides a novel route to tune the carrier concentration through both selective oxidation and electric gating. This establishes AV$_3$Sb$_5$  as a tunable 2D platform for the further exploration of topology and correlation among 3$d$ electrons in kagome lattices.

\end{abstract}
\maketitle
The kagome lattice, with an in-plane network of corner-sharing triangles, represents an ideal playground to realize various intriguing correlated phenomena\cite{1}, including flat-band ferromagnetism \cite{2,3}, spin liquids\cite{4}, charge density waves (CDW)\cite{5,6,7} and unconventional superconductivity\cite{5,6,7,8,9}. Additionally, exotic topological phenomena, such as Dirac and Weyl nodes, emerge in transition-metal kagome materials in combination with correlation effects\cite{Ye2018,LiuE2018,Morali2019,LiuDF2019,Yin2020}. Recently, a new family of vanadium-based kagome metals AV$_3$Sb$_5$ (A = K, Rb, Cs) have been discovered \cite{11}. Their electronic structure is characterized by a Z$_2$ topological invariant \cite{12} and superconductivity has been achieved with a maximum $T_{\mathrm{c}}$ of about 2.5 K in bulk samples at ambient pressure\cite{13,14}. Meanwhile, an intriguing CDW order\cite{15,16,17,18} is observed at higher temperature with possible time-reversal symmetry breaking and a giant anomalous Hall effect\cite{19,20}. Further investigations about superconductivity identified a pressure-induced double superconducting dome\cite{21,22,23}, a significant residual term in thermal conductivity\cite{24} and a V-shaped pairing gap in scanning tunneling microscopy measurements\cite{25,26}, hinting at a possible unconventional pairing symmetry with gap nodes. In contrast, penetration depth and nuclear magnetic resonance measurements suggest a nodeless gap\cite{duan2021nodeless,Mu}.

From the perspective of electronic structure, the Fermi level of  AV$_3$Sb$_5$ lies in the vicinity of multiple van Hove singularities (VHSs)\cite{27,28}, which are likely responsible for both CDW\cite{Feng,Denner,Lin} and superconductivity \cite{29}. Two of them are slightly below the Fermi level ($E_\text{F}$), with one of which having higher-order nature with a particularly flat dispersion \cite{hu}. Therefore, it is anticipated that introducing hole doping may shift those saddle points towards the Fermi level, providing an effective way of tuning Fermi-surface instabilities. However, previous studies on  AV$_3$Sb$_5$ have focused on bulk samples without carrier doping. Intentional charge modulation of the bulk sample through chemical doping has not been realized up to now. However, due to the quasi-2D nature and realtive weak interlayer interaction, it is possible to exfoliate thin flakes, for which carrier doping is more easily achievable. Moreover, in thin flakes quantum fluctuations and correlations are expected to be enhanced due to the reduced dimensionality. This may alter both supercondcutivity and the CDW, which in bulk samples has a 3D character with a nonzero $c$-axis modulation.
\begin{figure*}[tp]
	\includegraphics[clip,width=14cm]{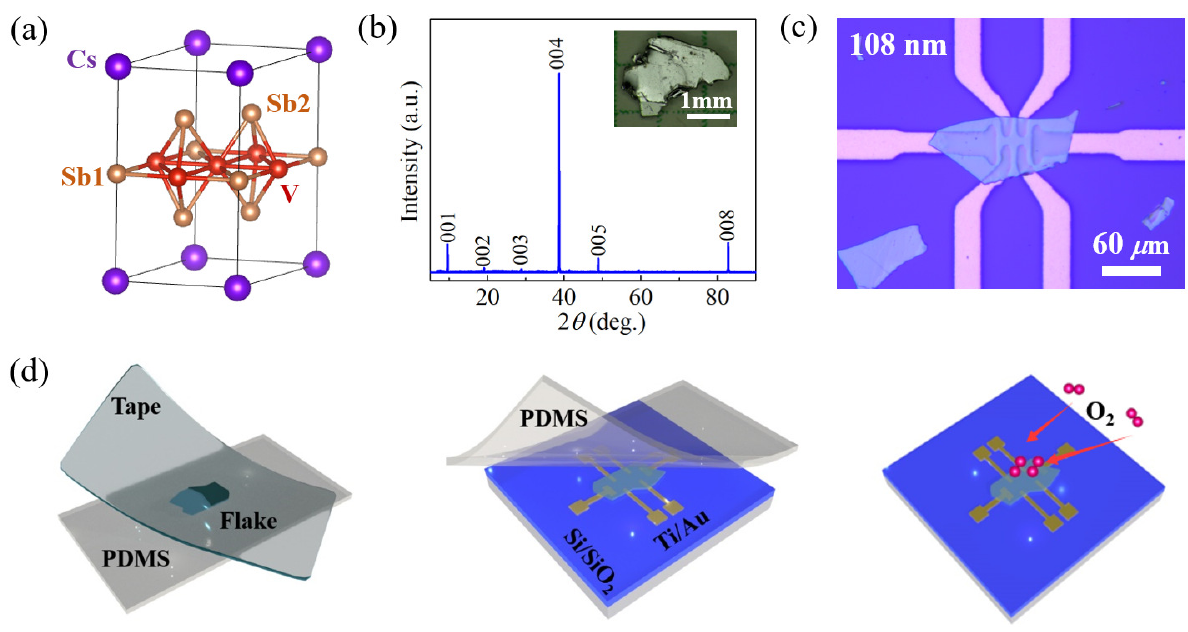}
	\caption{\label{fig1} Selective oxidation of CsV$_3$Sb$_5$ thin flakes. (a) Side view of the crystal structure of CsV$_3$Sb$_5$. (b) Room-temperature X-ray diffraction pattern of the CsV$_3$Sb$_5$ single crystal with a preferred orientation along the [00l] direction. The acquired bulk crystal shown in the inset. (c) Optical image of a typical fabricated device. (d) Illustration of the dry-transfer and selective-oxidation processes.}
\end{figure*}

In this Letter, we study the CDW and superconductivty in thin flakes of CsV$_3$Sb$_5$ and propose a novel doping method by selective oxidation, taking advantage of the semi-metallic nature of Sb. X-ray photoelectron spectroscopy (XPS) and gate-tuning experiments substantiate that hole doping in the topological kagome metals AV$_3$Sb$_5$ is realized for the first time. The $T_{\mathrm{c}}$ increases to 4.8 K with much enhanced and highly anisotropic upper critical fields, while the CDW is suppressed. A characteristic superconducting dome is revealed when considering the nonlinear relationship between the flake thickness and the effective doping. The $T_{\mathrm{c}}$ evolution follows the density of states at $E_\text{F}$, with a maximum when the higher-order VHS coincides with $E_\text{F}$, as modelled by our density functional theory (DFT) calculations. Our research provides a novel route to tune the correlated quantum states in layered kagome materials.

\begin{figure}
	\includegraphics[clip,width=8.5cm]{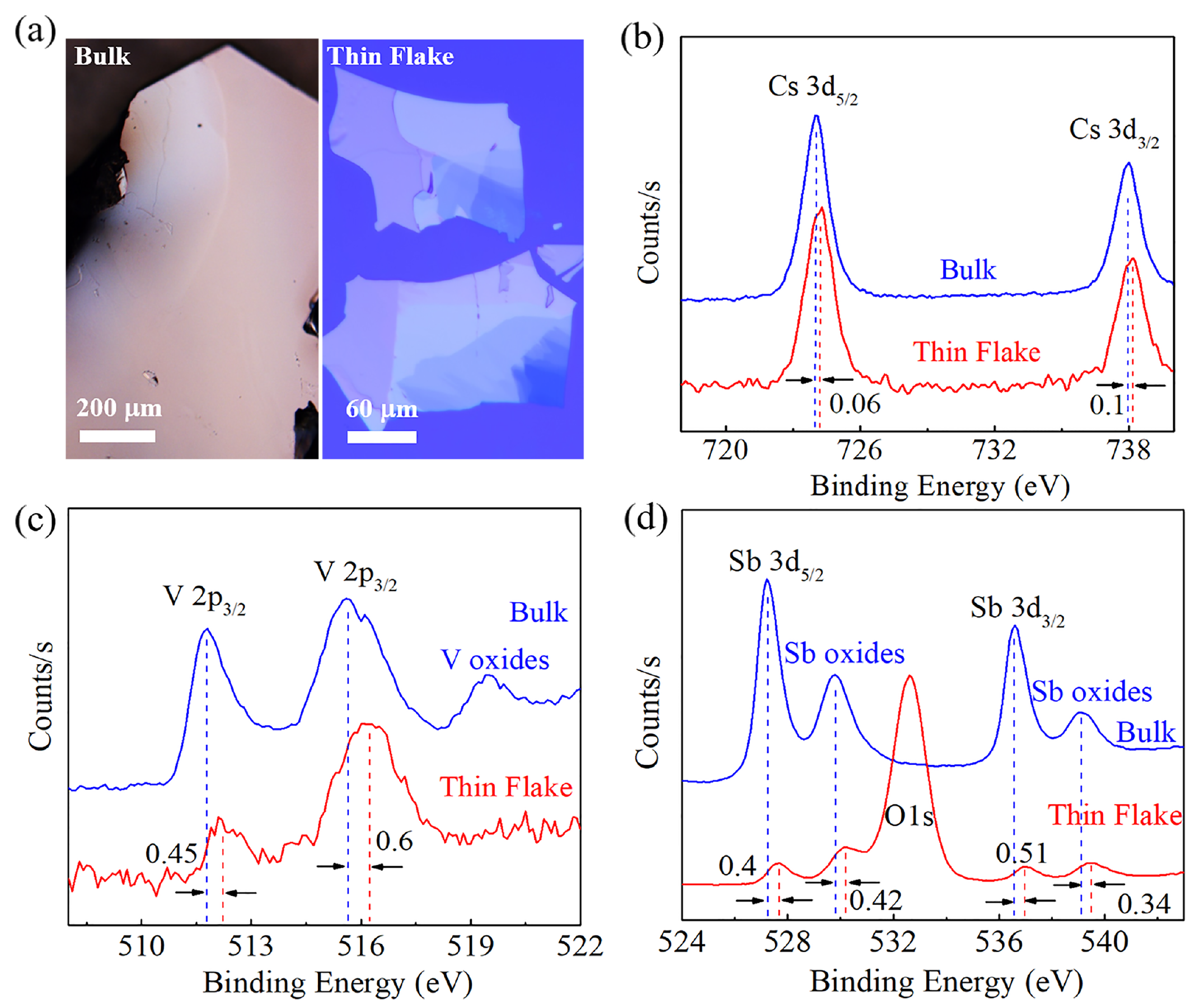}
	\caption{\label{fig2}X-ray photoelectron spectroscopy (XPS) measurements for bulk and thin flakes. (a) The optical images of bulk (left) and the exfoliated CsV$_3$Sb$_5$ thin flakes on a SiO$_2$/Si substrate (right). (b-d) The high-resolution XPS spectra of Cs 3${d}$, V 2${p}$, and Sb 3${d}$ for bulk crystal and thin flakes. The additional peak at 532.9 eV shown in (d) comes from the substrate.}
\end{figure}

The crystal structure of CsV$_3$Sb$_5$ shows $P$6/${mmm}$ symmetry with alternating stacking of kagome layers (V$_3$Sb$_5$) and charge reservoir layers (Cs), as displayed in Fig.\ref{fig1}(a). Our initial attempts of hole doping, such as substitution or decreasing the nominal Cs content, are not successful. However, we notice that the semi-metallic nature of antimony endows AV$_3$Sb$_5$ with moderate A-Sb interlayer interactions, which is much weaker than that in Co$_3$Sn$_2$S$_2$\cite{30} and provides a possibility to exfoliate  CsV$_3$Sb$_5$ into thin flakes. On the other hand, as the highest reactive metal, the surface Cs layer should be easily oxidized once exposed to air. We thus suggest a non-equilibrium route of selective oxidation to achieve hole doping in the intact  CsV$_3$Sb$_5$. The bulk single crystal (Fig.\ref{fig1}(b)) is first cleaved by using Scotch tape, and then the obtained thin flakes are transferred to pre-patterned electrodes (Fig.\ref{fig1}(c) and Fig. S1). The dry transfer method (Polydimethylsiloxane, PDMS) is adopted to avoid possible contaminations from the photoresist and other organic solutions. Before we perform measurements, the fabricated device is intentionally exposed to air for a few minutes for oxidation. Sample quality of the exfoliated thin flakes is characterized by using Raman spectroscopy (Fig. S2). Through controlling the thickness of the flakes, we could qualitatively modulate the carrier concentration.

\begin{figure}[tp]
	\includegraphics[clip,width=8.8cm]{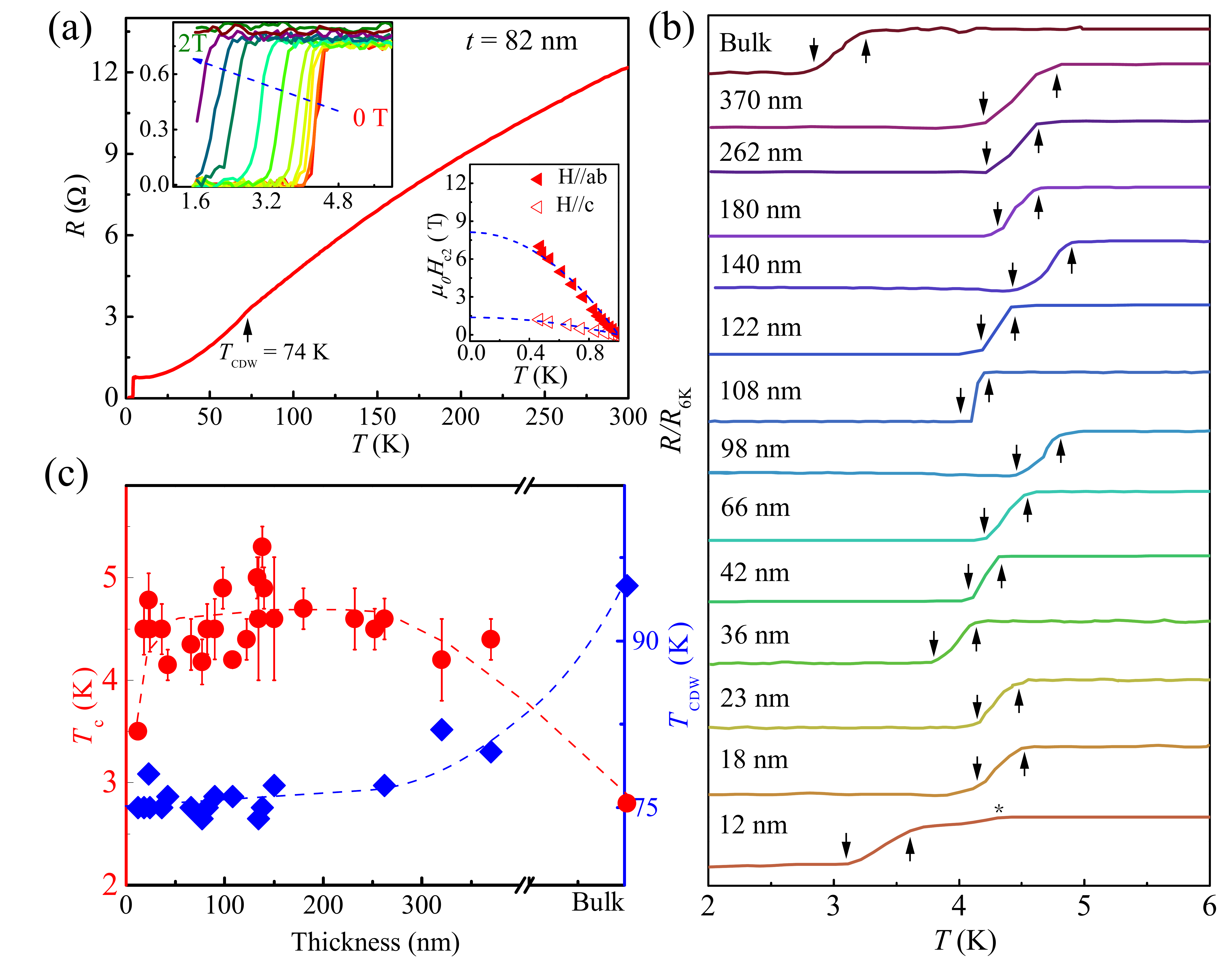}

	\caption{\label{fig3}Low-temperature transport property of CsV$_3$Sb$_5$ thin-flakes. (a) Temperature-dependent resistance of a 82 nm CsV$_3$Sb$_5$ thin fake. Upper inset shows the field-dependent ${R}$(T) curves at low temperatures. Lower inset is the upper critical field $H_{\mathrm{c2}}$ along the ${c}$-axis and ${ab}$ plane (raw data from which $H_{\mathrm{c2}}^{ab}$ extracted can be found in Fig. S4).  The solid lines are the Ginzburg-Landau fitting curves. (b) Temperature-dependent resistance for CsV$_3$Sb$_5$ thin fakes with various thicknesses. Each curve is vertically shifted to show the $T_{\mathrm{c}}^{onset}$ and $T_{\mathrm{c}}^{zero}$ more clearly. The $T_{\mathrm{CDW}}$  is determined by d$\rho$/dT and the raw data can be found in Fig. S5 and Fig. S6. (c) Temperature-thickness phase diagram of CsV$_3$Sb$_5$. Rhombus and circle symbols denote the $T_{\mathrm{CDW}}$ and $T_{\mathrm{c}}$, respectively. We use $T_{\mathrm{c}}^{mid}$, where the resistance drops 50\% of its normal state, as $T_{\mathrm{c}}$, and the transition widths, $T_{\mathrm{c}}^{onset}$ - $T_{\mathrm{c}}^{zero}$, are taken as the error bar. }
\end{figure}

To verify the possibility of hole doping, we compare the XPS of the bulk sample and the exfoliated thin flakes (Fig.\ref{fig2}(a)). The bulk sample was freshly cleaved before measurement to reduce the surface oxidation. As shown in Fig.\ref{fig2}(b-d), the peak shifts of Cs, V and Sb are 0.06, 0.45, and 0.5 eV, respectively. The negligible peak shift of Cs is due to the complete donation of Cs's 6$s^1$ electron to V$_3$Sb$_5$ layers. The apparent peak shifts towards higher binding energy in both V and Sb indicate the realization of hole doping in the cleaved thin flakes. It is worth pointing out that the oxidation peaks of V (515.6 to 516.2 eV) and Sb (529.7 to 530.12 eV) can be observed, implying that the exposed surface is terminated by the absorbed oxygen. Another independent evidence of hole doping can be given by gate-tuning experiment, which we will discuss below.

The achieved hole doping in the thin flakes allows us to investigate its influence on superconductivity and CDW orders. Fig. \ref{fig3}(a) shows the temperature-dependent resistance on a fabricated device with a thickness of 82 nm. Remarkably, its $T_{\mathrm{c}}^{onset}$ reaches 4.7 K, much higher than the bulk value at around 3 K (Fig. S3). Simultaneously, the CDW ordering temperature ($T_{\mathrm{CDW}}$) is reduced from 95 K to 74 K, coexisting with the superconductivity at low temperatures. The upper inset shows the magnetic-field suppression of the superconductivity. The extracted data are summarized in the lower inset, where the upper critical field acquired by fitting the Ginzburg-Landau formula is 1.38 T(8.31 T) for $H_c$ ($H_{ab}$), six (two) times higher than the bulk value of 0.24 T (4 T)\cite{12,32}.

\begin{figure}
	\includegraphics[clip,width=8.5cm]{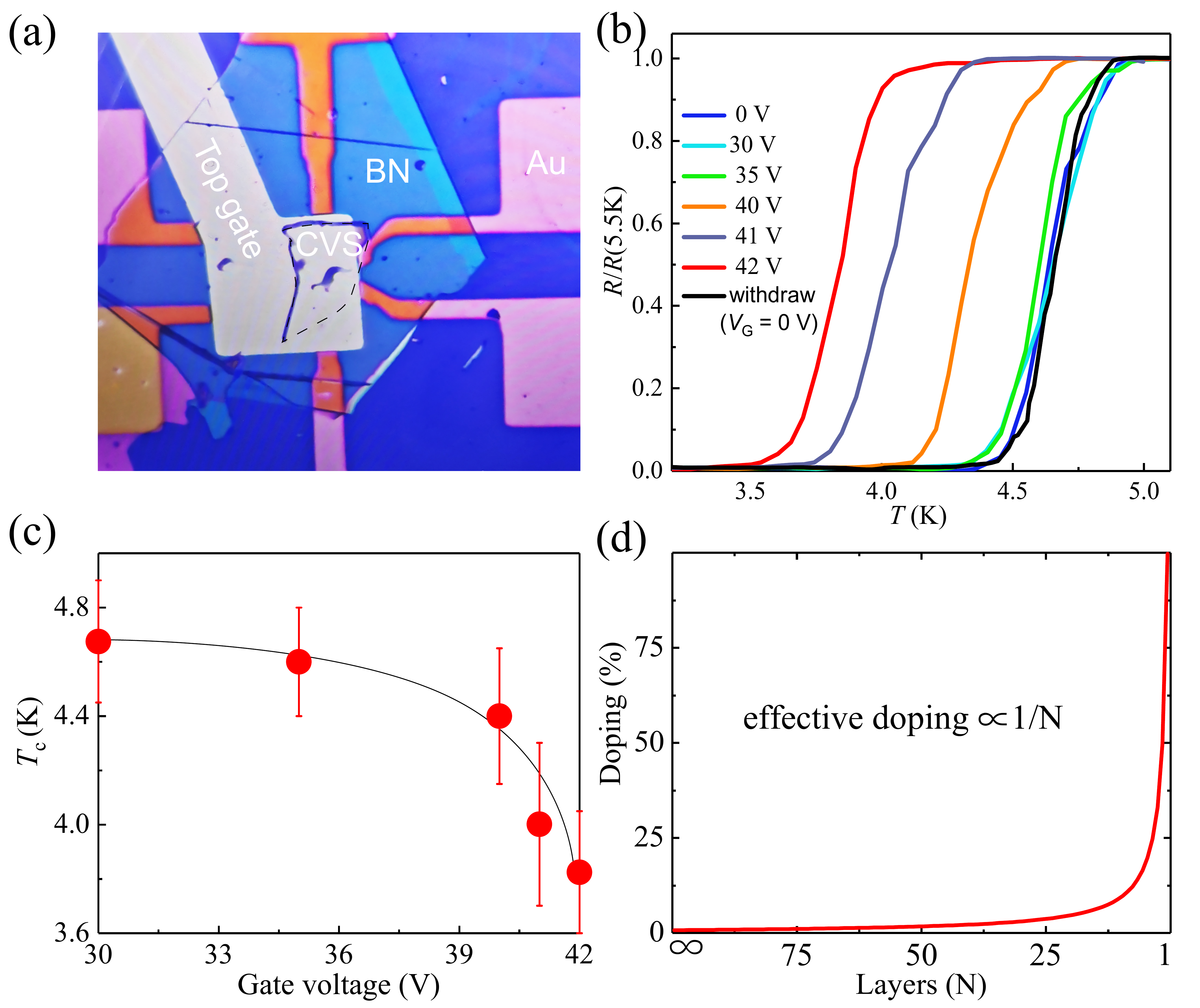}
	\caption{\label{fig4} Gate-tuning investigation of the selectively oxidized CsV$_3$Sb$_5$ thin flake. (a) Optical image of the fabricated device. The thickness of the sample and the BN gate insulator are 160 nm and 120 nm, respectively. The CsV$_3$Sb$_5$ thin flake was intentionally oxidized before the device fabrication. Positive gate voltage was applied on the top gate. (b) Temperature-dependent resistance under various gate voltages, with the results summarized in (c). (d) The non-linear relationship of flake thickness versus effective doping in our toy model estimation. The surface contribution will dramatically increase approaching the 2D limit.}
\end{figure}

\begin{figure*}
	\includegraphics[clip,width=16cm]{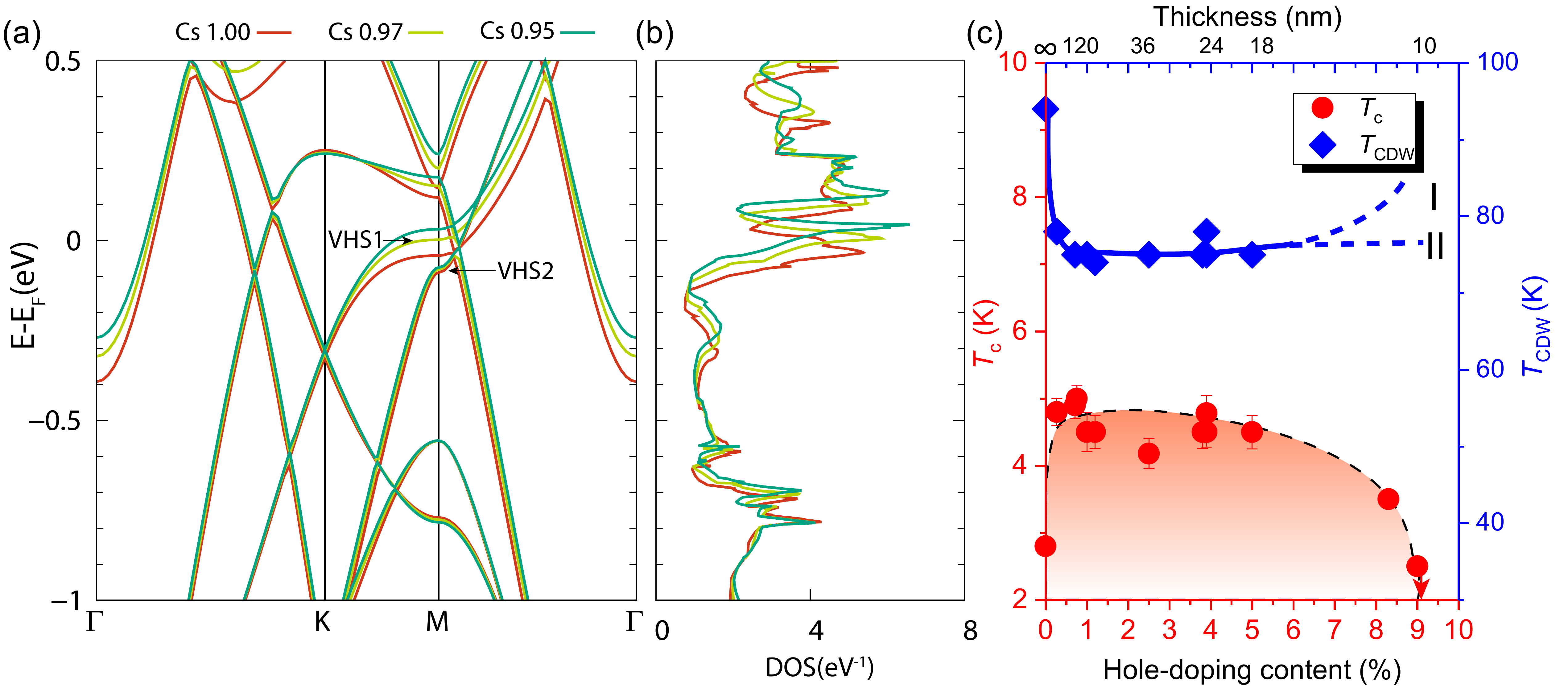}
	\caption{\label{fig5} Band structure, density of states (DOS), and phase diagram of hole doped CsV$_3$Sb$_5$. (a) Electronic band structure and (b) V $d$-orbital DOS for Cs$_x$V$_3$Sb$_5$ (x=1, 0.97, 0.95). (c) Phase diagram of CsV$_3$Sb$_5$ with the variation of hole-doping content, where only the top layer is assumed to be oxidized. The phase diagram in the low doping region ${(0 \thicksim 1\%)}$ can be found in Fig. S9, where fourteen devices with the flake thickness above 100 nm are included.}
\end{figure*}

We further investigate the thickness-dependent superconductivity and CDW transition temperatures, $T_{\mathrm{c}}$ and $T_{\mathrm{CDW}}$. A series of thin flakes from 12 nm to 370 nm are prepared and their typical temperature-dependent resistance are displayed in Fig.\ref{fig3}(b), where the arrow denotes the $T_{\mathrm{c}}^{onset}$ and $T_{\mathrm{c}}^{zero}$. The $T_{\mathrm{CDW}}$ is determined by standard d$\rho$/dT  procedure as shown in Fig. S5 and Fig. S6. The significant enhancement of onset $T_{\mathrm{c}}$ in thin flakes is clearly identified compared with bulk. A higher hole doping levels can be realized by using  thinner samples (e.g. 12 and 10 nm), where the $T_{\mathrm{c}}$ is surprise from 4.5 K to 3.5 and 2.5 K, respectively. The kink at 4.5 K (star in Fig. 3(b)) for the 12 nm sample is the superconducting signal arising from a thicker area due to the inhomogeneity of the exfoliated flake. Summarizing these results, we plot the temperature-thickness phase diagram of CsV$_3$Sb$_5$ in Fig.\ref{fig3}(c). The inverse correlation of $T_{\mathrm{c}}$ and $T_{\mathrm{CDW}}$ from bulk to thin flakes clearly demonstrates the inherent competing nature between superconductivity and CDW order. A similar phenomenon has been reported in cuprates\cite{nature}, heavy fermion compounds\cite{prx}, transition metal dichalcogenides\cite{TMD}, and documented in $\mathrm{Lu_5Ir_4Si_{10}}$ in a most impressive way\cite{2017}. With the application of external pressure or proton irradiation, $\mathrm{Lu_5Ir_4Si_{10}}$ exhibits clear evidence of the suppression of the CDW ordering and the enhancement of superconductivity\cite{1986,2020}.

By using the enhanced $T_{\mathrm{c}}$  as an indicator, it is possible to further provide a complementary evidence of the hole doping by conventional gate-tuning experiment. We expect that introducing electron doping in pre-oxidized samples can drive the $T_{\mathrm{c}}$ towards the bulk value, i.e., lower $T_{\mathrm{c}}$. As shown in Fig 4(a) and Fig. S7, a 160 nm CsV$_3$Sb$_5$ flake was firstly exposed to oxygen for a few seconds and capsulated by using a piece of 120 nm BN thin flake as the gate insulator. Electrons will accumulate on the sample side by applying a positive gating voltage at the top gate, which induces electron doping into the samples. With increasing the gating voltage, we find that the $T_{\mathrm{c}}$  continuously shifts to a lower temperature, and reaches 3.5 K at $V_{\mathrm{G}}$ = 42 V (the specific gating voltage depends on the thickness of the BN insulating layer). After retracting the gate voltage, the $T_{\mathrm{c}}$  quickly recovered to 4.8 K, indicating a pure gating effect (cyan line shown in Fig. 4(b)). The combination of XPS and gate-tuning results provide solid evidence of the realization of hole doping in CsV$_3$Sb$_5$ and the charge carrier-controlled $T_{\mathrm{c}}$ evolution. We note that this is the first time realizing the gating effect in the CsV$_3$Sb$_5$ material and our claim is also supported by a recent time-dependent ARPES measurements \cite{37}.

The variation of flake thickness induces a change of hole doping and hence has a prominent effect on the CDW and superconductivity. We further study the relationship between hole doping and flake thickness. When the thin flakes are exposed to air, the surface Cs layers are vulnerable towards oxidation. In order to estimate the doping content, we assume that only the top Cs layer is completely oxidized in flakes\cite{REF}. The effective hole doping is estimated to be proportional to $1/N$ with $N$ being the number of V$_3$Sb$_5$ kagome layers, as shown in Fig.\ref{fig4}(d). We find that the hole doping in thin flakes is prominent but weak in the bulk. With the layer separation being $c\approx$ 1 nm, the hole doping in the flakes with thickness ranging from 370 to 10 nm is about 0.2 to 10\%, respectively . For even thinner flakes, the top layer has a dramatic influence on the electronic structure and our estimation may no longer be valid.


To theoretically simulate it, we performed calculations with Cs vacancies using virtual crystal approximation. The band structures for the pristine and structures with 3\%, 5\% Cs vacancies are displayed in Fig.\ref{fig5}(a). Instead of a rigid-band shift, we find that Cs vacancies introduce noticeable hole doping mainly for the bands around $\varGamma$ and $M$ points. The band around the $M$ point is mainly attributed to Sb $p_z$ orbital and the VHS bands around the $M$ point are dominantly contributed by V $d$ orbitals (the VHS1 band is attributed to V  $d_{x^2-y^2 }$ and  $d_{z^2}$ orbitals, as illustrated in Fig. S8). Therefore, this orbital-selective hole doping by Cs vacancies is achieved through hybridization with Sb $p_z$ orbitals, inducing a prominent doping for Sb $p_z$-orbital and V $d_{z^2}$-orbital bands.

 Upon hole doping, the higher-order VHS1 point moves towards the Fermi level, resulting in an increase in DOS at $E_\text{F}$, as shown in Fig.\ref{fig5}(b). In particular, VHS1 touches the Fermi level with 3\% Cs vacancies and the $D(E_\text{F})$ reaches the maximum. Further increasing Cs vacancies, the hole-like Fermi surface around $\varGamma$ point undergoes a Lifshitz transition to electron-like Fermi surfaces centered around the $K$ points, generating a decreasing $D(E_\text{F})$. This evolution of $D(E_\text{F})$ may result in a superconducting dome independent of the pairing mechanism in AV$_3$Sb$_5$, similar to cuprates and iron-based superconductors\cite{33,34}. In contrast to the hexagonal Fermi surface from the conventional VHS featuring pronounced Fermi surface nesting\cite{nesting}, the higher-order VHSs exhibit flat dispersion along $K$-$M$ direction and a circular Fermi surface and thus has much weaker Fermi surface nesting. Therefore, $E_\text{F}$ approaching VHS1 will not promote CDW order. Based on this scenario of hole doping, we provide a qualitative explanation for the $T_{\mathrm{c}}$ enhancement and suppression of CDW order in thin flakes and we could understand why the bulk sample does not exhibit a higher $T_{\mathrm{c}}$ due to the negligible contribution from the surface.

Based on above analysis, the phase diagram can be drawn in terms of temperature versus effective hole doping, as shown in Fig.\ref{fig5} (c). A broad superconducting dome is evident,  although the hole doping levels may differ from the real composition. Since the $T_{\mathrm{CDW}}$ becomes almost undistinguishable in our 10 nm and 12 nm sample, we provide two possible scenarios (\Rmnum{1} and \Rmnum{2} in Fig.\ref{fig5} (c)) of the $T_{\mathrm{CDW}}$ at higher doping region. Low-doping region of the phase diagram of CsV$_3$Sb$_5$ is shown in Fig. S9. The gray area denotes the critical region from 370 nm to 70 $\mu$m, where the dramatic $T_{\mathrm{c}}$ enhancement takes place. Our detailed calculation also provides a similar dome-like behavior of the doping dependent DOS variation (shown in Fig. S10). To fully answer this question, thinner flakes below 10 nm are required, which is beyond the scope of the present paper and worth further investigation.

 There are three VHSs close to the Fermi level in AV$_3$Sb$_5$, two of which have sublattice-pure and one sublattice-mixed features \cite{29}. Moreover, VHS1 is of higher-order nature and VHS2 is conventional and features pronounced Fermi surface nesting. The orbital-selective doping from Cs vacancies has been shown to tune VHS1 but not VHS2. To study the effect of VHS2 on the correlated states, we can further introduce global-carrier doping in these thin flakes by thorough electrical gating and intercalation methods. This can be helpful to reveal the effects of different types of VHSs in promoting correlated phenomena in these kagome metals. The reduction of $T_{\mathrm{CDW}}$ in thin flakes suggests that CDW can still exit in quasi two dimensions. It would be interesting to further study its stability in the 2D monolayer limit. Experimentally, monolayer/bilayers can be realized by fine control of the oxidation. Strikingly, there is a significant change in the electronic structure of monolayer V$_3$Sb$_5$ (Fig. S11) according to our calculations: all VHSs are above $E_\text{F}$ and topological Dirac cones around the $K$ point are close to $E_\text{F}$. The combination of topological and correlation effects in the monolayer V$_3$Sb$_5$, which is tunable by electrical gating, may bring exotic quantum phenomena.

In conclusion, we realized the hole doping of the newly discovered topological kagome metals through a synergetic effect of thickness controlling and selective oxidation. Compared with bulk CsV$_3$Sb$_5$, the hole-doped thin flake shows a much higher $T_{\mathrm{c}}$ of about 4.8 K and a greatly enhanced $H_{\mathrm{c2}}$. The competing nature between superconductivity and CDW order is established and a comprehensive phase diagram is revealed, analogous to cuprates and iron-based superconductors. Band structure analysis demonstrates the crucial role of VHSs in the $T_{\mathrm{c}}$ enhancement with hole doping. The results presented here provide  new insight for the interpreation of superconductivity and CDW order in kagome systems. The proposed method may open the door to explore the kagome materials in the 2D limit, and the successful realization of gating effect can be applied to further reveal exotic correlation phenomena in these rich physical systems.\\

\noindent
${Note}$. We recently became aware of two independent works on thin flakes exfoliation for KV$_3$Sb$_5$\cite{35} and CsV$_3$Sb$_5$\cite{36}, where the $T_{\mathrm{c}}$ enhancement compared to that of the bulk is consistent with our paper. The proposed orbital-selective doping from the surface oxidation in present paper has been strongly supported by a recent ARPES paper\cite{37}.\\

\noindent
$^\dagger$ These authors contribute equally to this work.\\
$^*$ xianxin.wu@fkf.mpg.de\\
$^*$ jgguo@iphy.ac.cn\\
$^*$ chenx29@iphy.ac.cn\\

\noindent
${Acknowledgments}$. The authors thank Zhaoxu Chen, Xiangru Cui and Yuxin Yang for experimental help and Dr.Tongxu Yu for fruitful discussions. This work is supported by the National Key Research and Development Program of China (Grant Nos. 2017YFA0304700, 2019YFA0308000, 2018YFE0202601, and 2016YFA0300600), the National Natural Science Foundation of China (Grant Nos. 51922105, 62022089 and 51772322), the Chinese Academy of Sciences (Grant No.QYZDJ-SSW-SLH013), the Beijing Natural Science Foundation (Grant No. Z200005), the MoSTStrategic International Cooperation in Science, Technology and Innovation Key Program (2018YFE0202601) \\


\begin{thebibliography}{99}
\bibitem{REF} We note that this assumption is only qualitative and several top layers can get oxidized in experiments,  leading to a much higher doping content.
\end{thebibliography}
\end{document}